# Influence of the QCD Analogue of the Inverse Compton Effect on the Transverse Momentum and Pseudorapidity Distributions of Secondary Particles in *pp* Collisions at $\sqrt{s} = 30$ GeV, 510 GeV, and 14 TeV

M. Alizada and M. Suleymanov
Baku State University

## Abstract

Within the framework of numerical simulations, this work investigates the influence of the QCD analogue of the inverse Compton effect (ICE) in the quark–gluon scattering process $qg \to qg$ on the transverse momentum $p_T$ and pseudorapidity $\eta$ distributions of secondary particles produced in proton–proton collisions at energies $\sqrt{s} = 30$ GeV, 510 GeV, and 14 TeV. In the present context, ICE refers to a class of parton-level kinematic configurations in which the incoming quark carries a larger fraction of energy than the gluon, in contrast to the complementary DCE regime. The simulations were performed using the PYTHIA 8.316 event generator.
It is shown that the relative contributions of ICE and DCE strongly depend on the collision energy. As the energy increases from $\sqrt{s} = 30$ GeV to $\sqrt{s} = 14$ TeV, the ICE contribution becomes comparable to or exceeds the DCE contribution over a broad $p_T$range.
The analysis of pseudorapidity distributions demonstrates that deviations of the ICE/DCE ratio from unity appear predominantly in the central region $|\eta| \approx 0$, corresponding to symmetric partonic configurations $x_1 \sim x_2$, whereas in peripheral regions the ratio approaches unity. The obtained results indicate that, with increasing collision energy, the contribution of ICE-like processes grows due to the enhanced role of gluon collisions in the small-$x$ region.

## 1.Introduction

In Ref. [1], an analysis of quark–gluon scattering $gq \to gq$ in proton–proton collisions at $\sqrt{s} =$ 13.6 TeV was carried out in order to investigate the influence of kinematic configurations analogous to the inverse Compton effect (ICE) on the transverse momentum spectra $p_T$ of secondary particles. In this context, ICE does not represent a new dynamical mechanism, but rather a specific class of kinematic configurations within the standard process $qg \to qg$, in which the incoming quark possesses higher energy than the gluon. For comparison, configurations are considered in which the incoming gluon carries higher energy than the incoming quark in the center-of-mass system (DCE [2]).

The motivation for this study is related to works [3], in which the possibility of parton acceleration in a dense nuclear medium, analogous to the inverse Compton scattering process $\gamma + e^- \to \gamma + e^-$, where a relativistic electron transfers energy to a photon [4], was discussed. In the present case, the analogy is qualitative in nature and is used to describe the redistribution of energy between partonic degrees of freedom. It is expected that the ICE mechanism may lead to an enhancement in the production of particles with large $p_T$, i.e., to a hardening of transverse momentum spectra.

The origin of ultra-high-energy cosmic rays (UHECR, $E \gtrsim 10^{18}$eV) remains one of the key problems of astrophysics [5]. One possible mechanism involves the acceleration of partons in a dense medium [3]. In this context, it is important to understand how such kinematic configurations manifest themselves already in proton–proton collisions, which may serve as a baseline system.

The results obtained in Ref. [1] demonstrated that accounting for the contribution of quark–gluon scattering processes in proton–proton collisions at Large Hadron Collider energies [6] does not lead to a substantial hardening of transverse momentum spectra $p_T$, although it is accompanied by a moderate increase in the overall normalization of inclusive particle yields. Such behavior is explained by the fact that the process $qg \to qg$ forms the standard kinematic structure of two-particle ($2 \to 2$) partonic scattering within leading-order quantum chromodynamics (QCD) [7]. In this case, the spectral shape is determined mainly by the $t$-channel enhancement of the scattering amplitude and by the structure of parton distribution functions (PDFs) in hadrons [8].

Within the framework of collinear factorization, the observable cross section in hadron–hadron collisions is written as

$$d\sigma = \sum_{i,j} f_i\,(x_1, Q^2)\, f_j(x_2, Q^2)\, d\hat{\sigma}_{ij},$$

where $f_i(x, Q^2)$ are the parton distribution functions describing the probability of finding a parton of type $i$carrying a momentum fraction $x$at the scale $Q^2$, while $d\hat{\sigma}_{ij}$ is the elementary partonic scattering cross section calculated within QCD [7]. In the present work, a fixed PDF set (CT14QED) [9] is used, whereas the dynamics of the elementary processes are fully determined by the theory and are not modified.

Changes in the process kinematics are achieved by varying the collision energy $\sqrt{s}$, which corresponds to probing different regions of the Bjorken variable $x$ [10]: at high energies, small-$x$ regions dominated by the gluon sea become important, whereas at lower energies valence quarks provide a significant contribution. In addition, the type of colliding particles ($pp$, $pA$, $AA$) determines the properties of the partonic medium through the corresponding PDFs.

In the present work, the influence of the proton–proton collision energy on the transverse momentum $p_T$and pseudorapidity $\eta = -\ln\, \tan\,(\theta/2)$, where $\theta$ is the particle polar angle relative to the beam axis, is investigated with separation of different kinematic configurations of the process $qg \to qg$ (ICE and DCE). The considered energies are $\sqrt{s} = 30$ GeV, 510 GeV, and 14 TeV. The aim of the work is to study how the relative contribution of ICE configurations in the process $qg \to qg$depends on collision energy and how this affects the distributions of secondary particles in $p_T$ and $\eta$.

The pseudorapidity $\eta$correlates with the asymmetry of momentum fractions $x_1$and $x_2$ in $2 \to 2$ kinematics: the central region ($\eta \approx 0$) corresponds to configurations with $x_1 \sim x_2$, whereas large $|\,\eta\,|$values are associated with asymmetric configurations in which one parton carries a small momentum fraction and the other a large one. This makes pseudorapidity distributions a useful tool for analyzing the contributions of different kinematic regimes.

## 2. Methodology

Numerical simulations were performed using the PYTHIA event generator version 8.316 [11]. Proton–proton collisions at energies $\sqrt{s} = 30$ GeV, 510 GeV, and 14 TeV were considered.

To isolate the contribution of quark–gluon scattering, other hard interaction processes, in particular $gg \to gg$ and $qq \to qq$, were switched off in the generator settings. Thus, the analysis was carried out within a simplified model including only the $qg \to qg$ channel.

The CT14QED PDF set [9] was used to describe the internal structure of protons. Initial- and final-state parton shower effects (ISR and FSR) were enabled by default. Hadronization was modeled within the Lund string model [12] implemented in PYTHIA.

For each energy, $5 \times 10^5$ events were generated for each of the two classes of kinematic configurations, classified according to the relative energies of the initial partons in the center-of-mass system. The total statistics amounted to $10^6$events for each collision energy.

The analysis algorithm included:

- identification of partons participating in the hard scattering;
- classification of events according to the kinematic configuration type (ICE and DCE);
- selection of stable final-state hadrons;
- construction of transverse momentum $p_T$and pseudorapidity $\eta$distributions.

The following kinematic ranges were considered: $p_T \in \frac{[0,10]\text{ GeV}}{c}$; $\eta \in [-3.5,3.5]$, $[-6.5,6.5]$, $[-10,10]$ for energies $\sqrt{s} = 30$ GeV, 510 GeV, and 14 TeV, respectively. These ranges correspond to the full phase space available within the employed event generator.

## 3. Results

Based on the performed simulations, distributions of secondary particles in transverse momentum $p_T$and pseudorapidity $\eta$were obtained. Figure 1 (left panels) presents the inclusive cross sections for secondary particle production $d\sigma/dp_T$in proton–proton collisions at energies $\sqrt{s} = 30$ GeV (upper panel), $\sqrt{s} = 510$ GeV (middle panel), and $\sqrt{s} = 14$ TeV (lower panel) as functions of $p_T$. The light points correspond to Compton scattering processes (DCE), whereas the dark points describe events in which the inverse Compton scattering mechanism (ICE) is realized. For quantitative analysis of the spectra, similarly to Ref. [1], the following ratio was considered:

$$\text{Ratio} = \frac{(d\sigma/dp_T)_{\text{ICE}}}{(d\sigma/dp_T)_{\text{DCE}}}. \qquad (1)$$

The obtained results show that, within statistical uncertainties, the Ratio demonstrates different behavior as a function of $p_T$at different collision energies $\sqrt{s}$. At $\sqrt{s} = 30$ GeV, the Ratio is nearly independent of $p_T$ and remains approximately equal to 1.0 in the region $p_T < 3$ GeV/$c$. At $p_T > 3$ GeV/$c$, a decrease of the Ratio with increasing $p_T$is observed down to values of approximately Ratio $\sim 0.5$.

At $\sqrt{s} = 510$ GeV, in the region $p_T < 7$ GeV/$c$, the Ratio is also close to unity and, within statistical uncertainties, practically independent of $p_T$. In the interval $7 < p_T < 10$ GeV/$c$, a slight decrease of the Ratio to values of approximately $\sim 0.9$ is observed.

At $\sqrt{s} = 14$TeV, the Ratio becomes larger than unity within statistical uncertainties over the entire investigated $p_T$range (see Ref. [1]).

Thus, a pronounced energy dependence of the Ratio is observed, indicating a change in the relative contributions of ICE and DCE processes to the formation of transverse momentum spectra. It is seen that the relative contribution of ICE configurations increases with collision energy.

This result may be interpreted within the framework of changes in the dominant Bjorken-$x$ regions. At high energies, the processes mainly occur in the small-$x$ region, where the role of gluon contributions increases and more symmetric configurations $x_1 \sim x_2$are realized. This leads to an enhancement of the relative contribution of the considered class of events. In contrast, at low energies, asymmetric configurations involving valence quarks play a significant role, which leads to suppression of ICE configurations in the large-$p_T$ region.

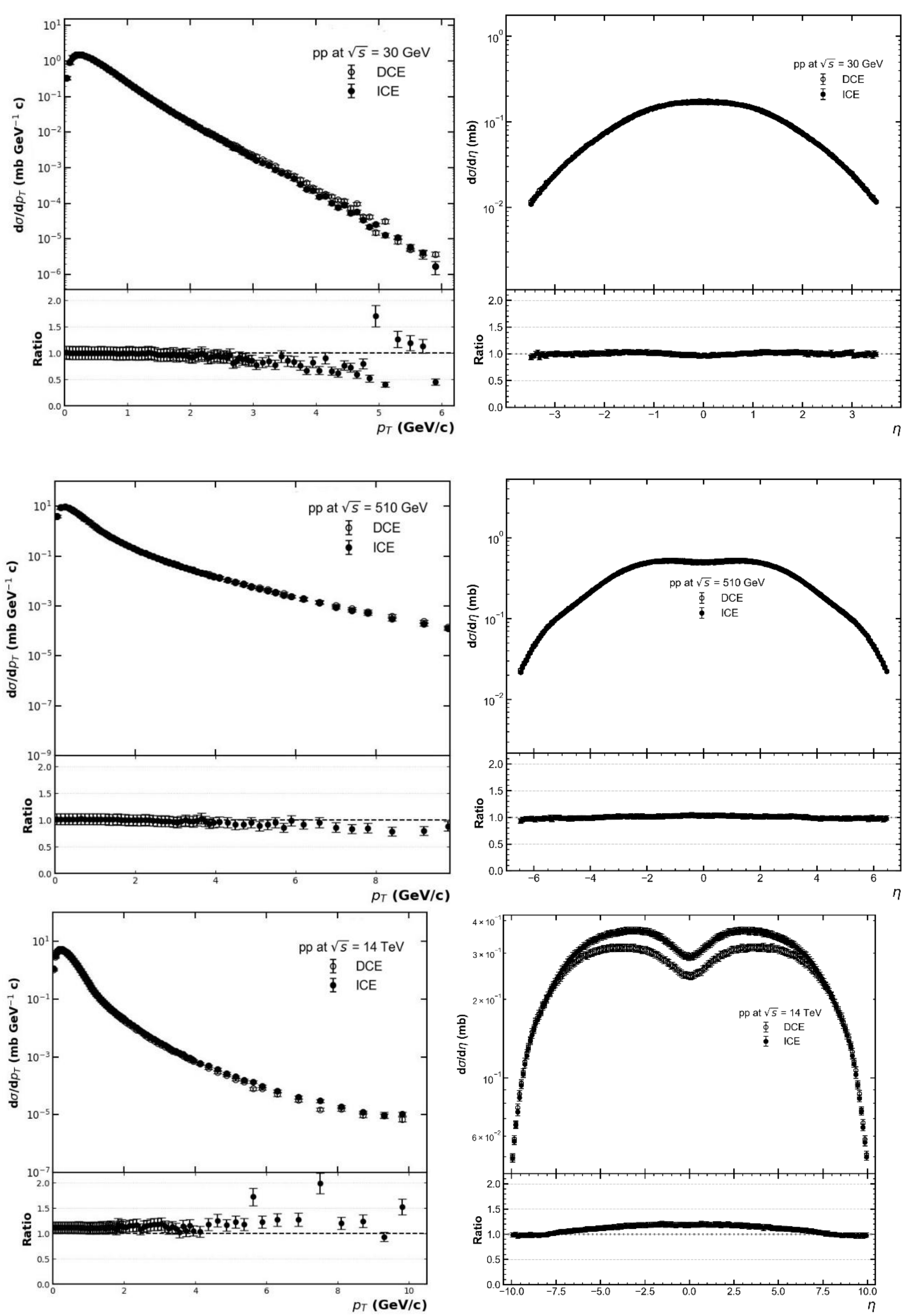

**Figure 1.** Inclusive cross sections for secondary particle production in proton–proton collisions at energies $\sqrt{s} = 30\text{GeV}$ (upper panels), $\sqrt{s} = 510\text{GeV}$ (middle panels), and $\sqrt{s} = 14\text{TeV}$ (lower panels). The left panels show transverse momentum distributions $p_T$, while the right panels present pseudorapidity distributions $\eta$.

The use of pseudorapidity $\eta$makes it possible to investigate the influence of the ICE mechanism in different kinematic regions. Since $\eta$is related to the parton momentum fractions $x_1$ and $x_2$, the analysis of $\eta$ distributions allow separation of the contributions from symmetric and asymmetric partonic configurations. In the central region ($\eta \approx 0$), conditions $x_1 \sim x_2$ are realized, whereas at large $|\eta|$values strong asymmetry emerges: one parton belongs to the small-$x$ region (gluon sea), while the other belongs to the large-$x$ region (valence quarks). This makes it possible to investigate the dependence of the ICE contribution on the kinematic regime and on the type of participating partons.

Figure 1 (right panels) presents the inclusive cross sections for secondary particle production $d\sigma/d\eta$in $pp$collisions at energies $\sqrt{s} = 30$GeV (upper panel), $\sqrt{s} = 510$GeV (middle panel), and $\sqrt{s} = 14$TeV (lower panel) as functions of pseudorapidity $\eta$. The spectra were obtained using the simulation procedure described above. The light points correspond to DCE configurations, while the dark points correspond to ICE configurations.

For quantitative analysis, the following ratio was considered:

$$\text{Ratio} = \frac{(d\sigma/d\eta)_{\text{ICE}}}{(d\sigma/d\eta)_{\text{DCE}}}. \qquad (2)$$

From Figure 1 (right panels), it can be seen that at energies $\sqrt{s} = 30$ GeV and 510 GeV, the Ratio is, within statistical uncertainties, nearly independent of $\eta$and remains close to unity. At $\sqrt{s} = 14$TeV, three characteristic regions can be distinguished in the Ratio dependence on pseudorapidity $\eta$:

$$\eta \in [-10.0, -7.5]\text{(Region I)},$$
$$\eta \in [-7.5, 7.5]\text{(Region II)},$$
$$\eta \in [7.5, 10.0]\text{(Region III)}.$$

In the peripheral regions I and III, the Ratio is approximately equal to unity, indicating a weak influence of the ICE mechanism in these regions. At the same time, in Region II ($-7.5 \leq \eta \leq 7.5$), which includes the central region ($\eta \approx 0$), a deviation of the Ratio from unity is observed. This indicates that the contribution of the ICE mechanism becomes significant specifically in the central pseudorapidity region corresponding predominantly to symmetric configurations $x_1 \sim x_2$.

At high energies, small values of the variable $x$are reached in this region, where the role of gluon processes increases, which may enhance the contribution of ICE configurations. In peripheral pseudorapidity regions ($|\eta| \geq 7.5$), corresponding to strongly asymmetric partonic configurations in which one parton belongs to the small-$x$ region and the other to the large-$x$ region, the contribution of the ICE mechanism remains insignificant.

Thus, the influence of the ICE mechanism on secondary particle spectra increases with collision energy. This effect is most pronounced in the small-$x$ kinematic region dominated by gluon processes, as well as in the region close to central pseudorapidity values.

## 4. Conclusions

In this work, the influence of two kinematic configurations of the process $qg \to qg$, namely the QCD analogue of the Compton effect (DCE) and the QCD analogue of the inverse Compton effect (ICE), on the transverse momentum and pseudorapidity distributions of secondary particles in proton–proton collisions at energies $\sqrt{s} = 30$ GeV, 510 GeV, and 14 TeV has been investigated within simulations performed using the PYTHIA 8.316 event generator.

It was found that the contribution of events classified as ICE configurations exhibits a pronounced dependence on collision energy:

- at $\sqrt{s} = 30$GeV, a noticeable suppression of the ratio

$$\text{Ratio} = \frac{(d\sigma/dp_T)_{\text{ICE}}}{(d\sigma/dp_T)_{\text{DCE}}}$$

is observed, reaching values of about 0.5in the large-$p_T$ region, indicating a lower probability of producing high-energy particles in such configurations. At the same time, the ratio

$$\frac{(d\sigma/d\eta)_{\text{ICE}}}{(d\sigma/d\eta)_{\text{DCE}}}$$

remains close to unity and practically independent of $\eta$;

- at $\sqrt{s} = 510$ GeV, the observed suppression becomes significantly weaker, and the Ratio decreases only to approximately $\sim 0.9$. The pseudorapidity distributions are also characterized by a ratio close to unity without pronounced dependence on $\eta$;
- at $\sqrt{s} = 14$ TeV, the ICE/DCE ratio becomes larger than unity (of the order of 1.1) over almost the entire investigated $p_T$range. In addition, a pronounced kinematic dependence of the ICE contribution on pseudorapidity $\eta$ is observed. In the central region ($|\eta| \leq 7.5$), corresponding predominantly to symmetric partonic configurations $x_1 \sim x_2$ and small values of the Bjorken variable $x$, a deviation of the ICE/DCE ratio from unity is observed. In peripheral regions ($|\eta| \geq 7.5$), corresponding to asymmetric “gluon–valence quark” configurations, the ICE contribution remains insignificant.

The observed energy dependence is related to changes in the dominant Bjorken-$x$ regions: at low energies, valence quarks (large $x$) play an essential role, whereas at high energies the gluon sea (small $x$) dominates, leading to more symmetric kinematic configurations of partonic interactions. The analysis of pseudorapidity distributions showed that the influence of ICE configurations is weakly manifested in integrated $\eta$-spectra at energies $\sqrt{s} = 30$ GeV and 510 GeV, whereas at $\sqrt{s} = 14$ TeV a noticeable deviation of the ICE/DCE ratio from unity is observed in the central region corresponding to configurations $x_1 \sim x_2$. The obtained results confirm that $pp$collisions may be considered as a baseline system for investigating energy redistribution effects at the parton level. At the same time, the QCD analogue of the inverse Compton effect by itself does not lead to substantial hardening of particle spectra, and its influence is determined mainly by the kinematic features of the process and by the structure of parton distribution functions.

## References

1. M. Alizada, M. Suleymanov. The influence of the inverse Compton effect on the transverse momentum spectra of particles produced in pp collisions at $\sqrt{s}$=14 TeV. hep-ph>arXiv:2604.18654
2. Fritzsch H and Minkowski P 1977 Phys. Lett. B 69 316–32
3. Suleymanov M 2016 Georgian Electronic Scientific Journal: Physics 1(15) 92. Suleymanov M 2009 Proceedings of Science EPS-HEP2009 406; Suleymanov M . 2012 J. Phys.: Conf. Ser. 347 012024
4. Kristy E. McGhee. A century of Compton scattering. Nature Reviews Physics.5,322(2023)
5. A.V. Olinto. Ultra-high energy cosmic rays: the theoretical challenge. Physics Reports Volumes 333–334, August 2000, Pages 329–348]
6. https://home.cern/science/accelerators/large-hadron-collider 109-1312 .
7. R. K. Ellis, W. J. Stirling, B. R. Webber. QCD and Collider Physics. Cambridge Monographs on Particle Physics, Nuclear Physics and Cosmology.Cambridge University Press. 05 May 2010
8. J F Owens and W Tung. Parton Distribution Functions of Hadrons. Annual Review of Nuclear and Particle Science. Vol. 42:291-332 (1992), https://doi.org/10.1146/annurev.ns.42.120192.001451
9. Carl Schmidt, Jon Pumplin, Daniel Stump, and C.-P. Yuan. CT14QED parton distribution functions from isolated photon production in deep inelastic scattering. Phys. Rev. D 93, 114015 (2016). DOI: https://doi.org/10.1103/PhysRevD.93.114015

10. J. D. Bjorken. Asymptotic Sum Rules at Infinite Momentum. Phys. Rev. 179, 1547 , 1969. DOI: https://doi.org/10.1103/PhysRev.179.1547
11. Sjöstrand T et al. 2015 Comput. Phys. Commun. 191 159–177
12. Bo Andersson et al.*Parton Fragmentation and String Dynamics*. Physics Reports 97 (1983) 31–145. DOI: https://doi.org/10.1016/0370-1573(83)90080-7